\documentstyle[prl,aps,twocolumn,epsfig]{revtex}

\begin{document}
\twocolumn[\hsize\textwidth\columnwidth\hsize\csname @twocolumnfalse\endcsname

\title{The monoclinic phase in PZT: new light on morphotropic phase boundaries}

\author{B. Noheda$^1$, J. A. Gonzalo} 
\address{Universidad Autonoma de Madrid, Cantoblanco 28049, Spain}
\author{R. Guo, S.-E. Park, L.E. Cross} 
\address{Materials Reasearch Lab., Penn. State University, PA 16802}
\author{D.E. Cox, and G. Shirane} 
\address{Physics department, Brookhaven National Lab., Upton,  NY 11973}
\maketitle

\begin{abstract}
A summary of the work recently carried out on the morphotropic phase
boundary (MPB) of PZT is presented. By means of x-ray powder diffraction on
ceramic samples of excellent quality, the MPB has been successfully
characterized by changing temperature in a series of closely spaced compositions. 
As a result, an
unexpected monoclinic phase has been found to exist in between the
well-known tetragonal and rhombohedral PZT phases. A detailed structural
analysis, together with the investigation of the field effect in this region
of compositions, have led to an important advance in understanding the
mechanisms responsible for the physical properties of PZT as well as other
piezoelectric materials with similar morphotropic phase boundaries.
\end{abstract}


\vskip1pc]
\narrowtext
\preprint{cond-mat/000-ms}
\newpage

\section*{Introduction}\footnote{To appear in the proceedings of the Workshop on 
{\it Fundamental Physics of Ferroelectrics} held in Aspen, Feb. 2000.

$^1$Present address: Physics department, Brookhaven National Lab., Upton,  NY 11973. }The morphotropic phase boundary (MPB) of PbZr$_{1-x}$Ti$_x$O$_3$ (PZT) has
been finally characterized on extremely homogeneous ceramic samples by high
resolution x-ray measurements. The boundary has been found to define the
limit between the tetragonal phase and a new PZT phase with monoclinic
symmetry. The recent work on this finding is reviewed \cite
{Noheda1,Noheda2,Noheda3,Noheda4}.

\begin{figure}[h]
\epsfig{width=0.80 \linewidth,figure=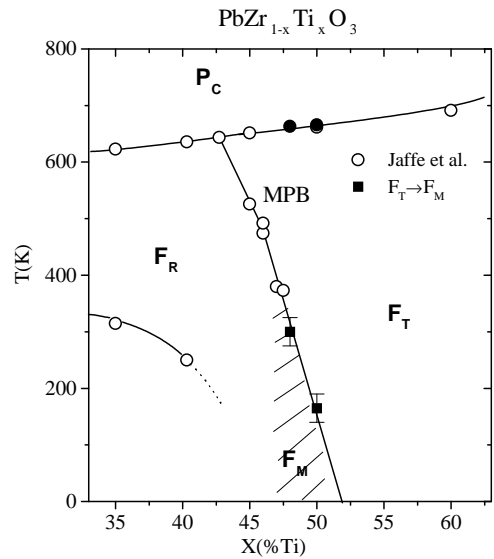}
\caption{Preliminary modification of the PZT phase diagram after Noheda et al.
\protect\cite{Noheda3}. The open symbols represent the PZT phase diagram
after Jaffe et al. \protect\cite{Jaffe}.}
\label{fig1}
\end{figure}

The remarkable physical properties of the ferroelectric system PbZr$_{1-x}$Ti%
$_{x}$O$_{3}$ (PZT) for compositions close to x= 0.47 have been known for
many years. In particular, its high piezoelectric response has made PZT one
of the most widely used materials for electromechanical applications.
PZT was first studied five decades ago \cite{Shirane,Sawaguchi} and the main
structural characteristics of the system were investigated at that time. At
high temperatures PZT is cubic with the perovskite structure. When lowering
the temperature the material becomes ferroelectric, with the symmetry of the
ferroelectric phase being tetragonal (F$_{T}$) for Ti-rich compositions and
rhombohedral (F$_{R}$) for Zr-rich compositions. Subsequent studies led to
the generally accepted phase diagram after Jaffe et al.\cite{Jaffe}, which
covers temperatures above 300 K. Jaffe's phase diagram is represented by
open circles in Fig. \ref{fig1} for 0.33$\leq $x $\leq $0.63. A complete
phenomenological theory was developed for this system that is able to
calculate thermal, elastic, dielectric and piezoelectric parameters of
ferroelectric single crystal states \cite{Haun}.

The boundary between the tetragonal and the rhombohedral phases, at
compositions close to x= 0.47, the so-called morphotropic phase boundary
(MPB)\cite{Jaffe}, is nearly vertical in temperature scale. It has been
experimentally observed that the maximum values of the dielectric
permittivity, as well as the electromechanical coupling factors and
piezoelectric coefficients of PZT at room temperature occur on this phase
boundary. However, the maximum value of the remanent polarization is shifted
to smaller Ti contents \cite{Jaffe}.

The space groups of the tetragonal and rhombohedral phases (P4mm and R3m,
respectively) are not symmetry-related, so a first order phase transition is
expected at the MPB. However, this has never been observed and, as far as we
know, only composition dependence studies are available in the literature.
One of the main difficulties in the experimental approach to this problem is
the lack of single crystals of PZT. Because of the steepness of the phase
boundary, any small compositional inhomogeneity leads to a region of phase
coexistence (see e.g.\cite{Kakewaga,Mishra,Zhang,Wilkinson}) that conceals
the tetragonal-to-rhombohedral phase transition. The width of the
coexistence region has been also connected to the particle size \cite{Cao}
and depends on the processing conditions, so a meaningful comparison of
available data in this region is often not possible.

On the other hand, the richness of phases in PZT and the simplicity of its
unit cell have encouraged important theoretical efforts in recent years. So
far, the first-principles studies have been successful in reproducing many
of the physical properties of PZT \cite{Saghi,Bellaiche}. But, in spite of
the proven validity, these calculations had not yet accounted for the
remarkable increment of the piezoelectric response observed when the
material approaches its MPB.

\begin{figure}[h]
\epsfig{width=0.95 \linewidth,figure=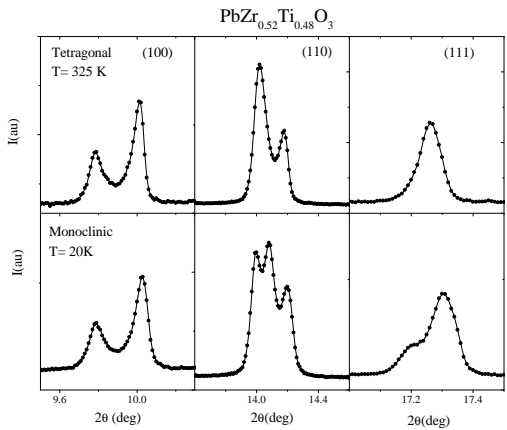}
\caption{Three selected peaks, (100), (110) and (111), from the diffraction
profiles in the tetragonal (top) and monoclinic (bottom) PZT (x= 0.48) phases. The diffraction 
profiles are shown in Noheda et al.\protect\cite{Noheda3}, Fig.6.}
\label{fig2}
\end{figure}

It was accordingly clear that there was something missing in the understanding of
PZT, mainly due to experimental difficulties, and we addressed our efforts
in this direction. The slight deviation from verticality of the MPB
encouraged us to attempt the investigation of a temperature driven $%
F_{T}-F_{R}$ phase transition knowing that only samples of exceptional
quality would allow us to succeed. With this purpose such ceramics were prepared
by the Penn. State group. Our collaboration has resulted in the discovery of
a new monoclinic phase (F$_{M}$) in this ferroelectric system \cite
{Noheda1,Noheda2}. The detailed structural analysis of tetragonal \cite
{Noheda3} and rhombohedral\cite{Corker} phases of PZT seemed to indicate
that the local structure is different from the average one and that, in both
phases, such local structure has monoclinic symmetry. This local
structure would be the precursor of the observed monoclinic phase.
Diffraction measurements of the effect of the electric field on ceramic
samples have confirmed this model \cite{Noheda4}. Measurements on PZT samples with x= 0.48
and x= 0.50 allowed for a modification of the PZT phase diagram as shown in
Fig. \ref{fig1}. It should be noted that the MPB defined by Jaffe et al. is
still a perfectly valid line that corresponds to the F$_{T}$-F$_{M}$ phase
transition.

\section*{Experimental}

 PZT samples were prepared by conventional solid-state reaction techniques as described in
refs.\cite{Noheda1,Noheda3} at the MRL, Penn. State University, and samples
with x= 0.50 were prepared at ICG, CSIC, Spain, as described in ref.\cite
{Noheda2}. High-resolution synchrotron x-ray powder diffraction measurements
were carried out at the X7A diffractometer, at the Brookhaven National
Synchrotron Light Source. Two types of experiments were done, as explained
in ref. \cite{Noheda3}. In the first one, data were collected from a disk in
a symmetric flat-plate reflection geometry over selected angular regions as
a function of temperature. These measurements demonstrated the high quality
of the ceramic samples, whose diffraction peaks in the cubic phase have
full-widths at half-maximum of 0.02$^{\circ }$ for x= 0.48. By means of a
Williamson-Hall analysis in the cubic phase, a compositional error of less
than $\Delta x=\pm 0.003$ was estimated \cite{Noheda3} for this composition.
To perform a detailed structure determination, additional measurements
were made for PZT with x= 0.48, at 20 and 325 K, in the monoclinic and
tetragonal phases, respectively. In this case, the sample was loaded in a
rotating capillary of 0.2 mm diameter to avoid texture and preferred
orientation effects \cite{Noheda3}.

\section*{The monoclinic phase}

According to the PZT phase diagram\cite{Jaffe}, a sample with x= 0.48 is
tetragonal just below the Curie point and rhombohedral below room
temperature. The measurements on the pellets for selected diffraction peaks,
with decreasing temperature from the cubic phase, showed the expected
tetragonal phase down to $\sim $300 K. Below this temperature however new
features appeared in the diffractograms, but they were not compatible
with either a rhombohedral phase or with a mixture of both phases (tetragonal and
rhombohedral), and they clearly corresponded to a monoclinic phase
with b as a unique axis \cite{Noheda1}. This can be
observed in Figure \ref{fig2} where selected parts of the diffraction
profile are plotted for the monoclinic (20 K) and the tetragonal (325 K)
phases.

\begin{figure}[h]
\epsfig{width=0.95 \linewidth,figure=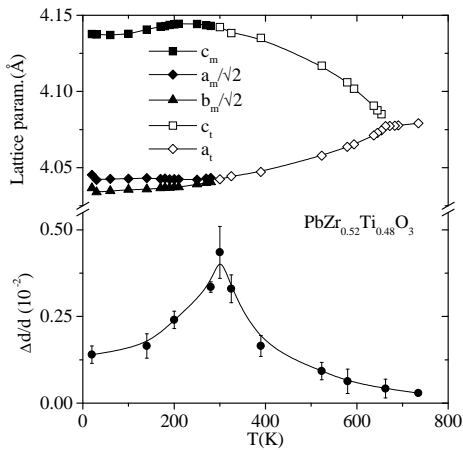}
\caption{Cell parameters as a function of temperature in the cubic,
tetragonal and monoclinic phases for PZT x= 0.48. The evolution with
temperature of $\Delta d/d$ is shown is the bottom plot. After Noheda et al.
\protect\cite{Noheda3}}
\label{fig3}
\end{figure}

The cell parameters of PZT (x=0.48) are represented in Fig. \ref{fig3} as a
function of temperature. In the tetragonal phase (below T$_{c}$= 660 K), the
tetragonal strain $c_{t}/a_{t}$, increases as the temperature decreases. At T%
$\simeq $ 300K the tetragonal-to-monoclinic ($F_{T}-F_{M}$) phase transition
takes place and the $c_{t}/a_{t}$ ratio starts decreasing slightly. The
microstrain present in the sample during the evolution of the tetragonal
phase seems to play a crucial role in the phase transition. $\Delta d/d$ is 
obtained from a Williamson-Hall analysis of the diffraction line widths\cite{Noheda3} 
and is shown as a function of temperature at the bottom of the plot. At high temperatures, 
$\Delta d/d$ increases as the temperature decreases. At first, the increment
is slow and no anomaly is observed at the cubic-to-tetragonal transition. At
lower temperatures $\Delta d/d$ shows a rapid increase that reaches a sharp maximum 
just at the $F_{T}-F_{M}$ phase transition. For this amount of Zr substituted, 
the tetragonal phase cannot support the stress in the structure, 
which is to a large extent released by the onset of the monoclinic phase. 
Further analysis is being done in order to compare the microstrain exhibited by 
different compositions with that
observed in pure PbTiO$_{3}$, where the tetragonal phase is stable at very
low temperatures.

\begin{figure}[h]
\epsfig{width=0.90\linewidth,figure=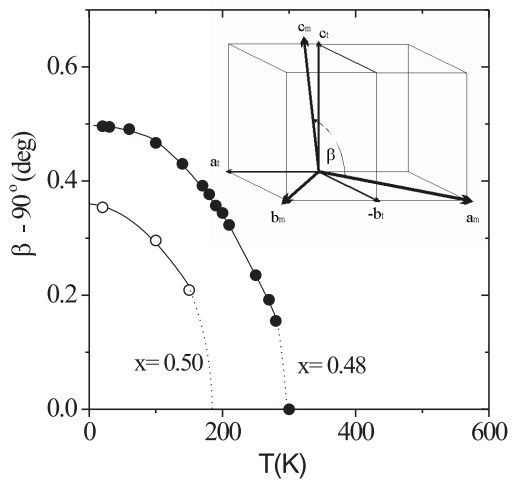}
\caption{Temperature evolution of the order parameter of the $F_T-F_M$
transition for PZT with x= 0.48 and x= 0.50. A representation of the
monoclinic unit cell is included as an inset. After Noheda et al.\protect\cite
{Noheda1,Noheda2,Noheda3}.}
\label{fig4}
\end{figure}

The monoclinic unit cell is doubled with respect to the tetragonal one: $%
a_{m}$ and $b_{m}$ are aligned along the [$\overline{1}\overline{1}0]$ and [1%
$\overline{1}0$] directions, respectively, and $c_{m}$ remains approximately
equal to the tetragonal $c_{t}$ but it is tilted with respect to it in the
monoclinic plane, as illustrated in the inset in Fig. \ref{fig4}. Such a unit
cell is chosen in order to have a monoclinic angle, $\beta $, larger than 90$%
^{\circ }$ (according to the usual convention) and $\beta -90^{\circ }$ is
then defined as the order parameter of the $F_{T}-F_{M}$ transition. Its
temperature evolution is depicted in Fig. \ref{fig4}. The transition seems
to be of second order which is allowed in this case, since Cm is a subgroup
of P4mm. PZT samples with x= 0.50, prepared in a slightly different way \cite
{Noheda2}, showed also a monoclinic phase for temperatures below 200 K. In
this case $a_{m}$ is approximately equal to $b_{m}$ and the monoclinic
angle, $\beta $ was found to be smaller than that observed for x= 0.48. Its
evolution with temperature is also plotted in Fig.\ref{fig4} \cite{Noheda2}.
A direct comparison of these data for samples from different origins must be regarded 
with caution. Further work is being carried out in which samples with compositions
in the range x= 0.42-0.51 processed under the same conditions are studied.
However, with the data obtained so far it is already possible to represent a
modification of the PZT phase diagram as the one shown in figure \ref{fig1}.
It can be observed that the MPB established by Jaffe et al.\cite{Jaffe} above
room temperature seems to lie exactly along the F$_{T}$-F$_{M}$ phase
boundary.

Contrary to what occurs in the tetragonal and rhombohedral phases, in the
monoclinic phase the polar axis is not determined by symmetry and could
be along any direction within the monoclinic plane. To determine this
direction the atom positions need to be known. A detailed structure
investigation by means of a Rietveld profile analysis of the tetragonal and
monoclinic phases of PZT (x= 0.48) has produced interesting results \cite
{Noheda3}. In the tetragonal phase at 325 K the unit cell has $a_{t}=4.044%
\AA $ and $c_{t}=4.138\AA $ and the atoms were found to be displaced in the
same way as in pure PbTiO$_{3}$: Pb and Zr/Ti were shifted 0.48 \AA\ and 0.27 
\AA , respectively, along the polar [001] axis. Anisotropic temperature
factors gave a much better refinement but the resultant thermal ellipsoids
were unphysically flattened perpendicularly to the polar direction. This is
not a new problem in PZT: Rietveld refinement of the rhombohedral PZT\cite
{Glazer} structure also produced thermal disk-shaped ellipsoids flattened
perpendicular to the rhombohedral polar axis [111]. This observed behavior
has been previously associated with the existence of certain local ordering
different from the long-range order.

\begin{figure}[h]
\epsfig{width=0.70 \linewidth,figure=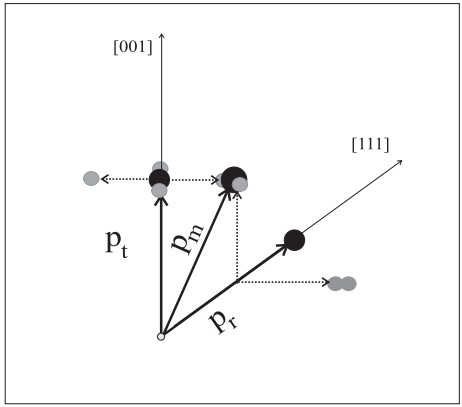}
\vskip2pc
\caption{Schematic view of the cation shifts in the tetragonal (p$_t$),
monoclinic (p$_m$) and rhombohedral ($p_r$) unit cells. The solid circles
and the arrows represent the cation positions in each one of the three
structures. In the tetragonal and rhombohedral cases, the long-range
structure is an average of the short-range or local structures (three in the
rhombohedral one and four in the tetragonal one) represented by smaller grey
circles \protect\cite{Noheda3}}
\label{fig5}
\end{figure}

The local order has been studied in rhombohedral PZT by means of the Pair
Distribution Function \cite{Teslic} and, more recently, by modelling local
disordered cation shifts by means of a Rietveld profile refinement \cite
{Corker}. The authors found that by considering three equivalent disordered
displacements along the $\langle $001$\rangle $ directions, superimposed on
the rhombohedral cation displacement along [111] (see figure \ref{fig5}) the
refinement produced much more reasonable temperature factors.

In the same way, for tetragonal PZT(x= 0.48) at 325 K, we can model
local disordered sites for the lead atoms perpendicular to the polar axis,
that is, we allow Pb to move towards the four sites allowed by symmetry,
i.e. at the $\langle $xxz$\rangle $ positions \cite{Noheda3}, which give an
average tetragonal (00z) cation shift (see figure \ref{fig5}). Similar
results are obtained if the refinement is carried out modelling local
disorder shifts along the $\langle $x0z$\rangle $ directions. The refinement
gives local shifts of ~0.2 \AA\ perpendicular to the polar axis in adition
to the common shift of 0.48 \AA\ along the polar axis, which is similar to
that of PbTiO$_{3}$, and gives also physically reasonable isotropic
temperature factors.

The structure of the monoclinic phase at 20 K does not present this kind of
problems. The refinement is very good considering isotropic temperature
factors for all atoms except lead, and the resulting anisotropy for lead is
not unreasonable \cite{Noheda3}. The refined unit cell was $a_{m}=5.721$ $%
\AA $, $b_{m}=5.708$ $\AA $, $c_{m}=4.138$ $\AA $ with $\beta =90.496^\circ$%
. The results os the refinement \cite{Noheda3} have defined the monoclinic
polar axis. This lies within the monoclinic plane along a direction between 
the polar axes of the tetragonal and the rhombohedral phases, 
$24^\circ$ away from the former (see figure \ref{fig5}). This value could
become slightly different after the oxygen positions are more accurately
determined by a neutron study that is underway. This is the first example
of a ferroelectric material with $P_{x}^{2}=P_{y}^{2}\neq P_{z}^{2}$, where $%
P_{x}$,$P_{y}$ and $P_{z}$ are the Cartesian components of the spontaneous
polarization.

Although this result is interesting, the striking fact about it is that the
monoclinic shifts exactly corresponds to one of the four locally disordered
shifts proposed for the tetragonal phase, as it can be observed in fig.\ref
{fig5}. The monoclinic phase appears, as the temperature is lowered, by the
condensation of one of the local shifts existing in the tetragonal phase.
Most interesting is the fact that the monoclinic displacement also
corresponds to one of the three locally disordered shifts proposed by Corker
et al.\cite{Corker} for the rhombohedral phase (see fig. \ref{fig5}), so the
condensation of this particular site would also give rise to the observed
monoclinic phase.

\section*{Field effect}

\begin{figure}[h]
\epsfig{width=0.90 \linewidth,figure=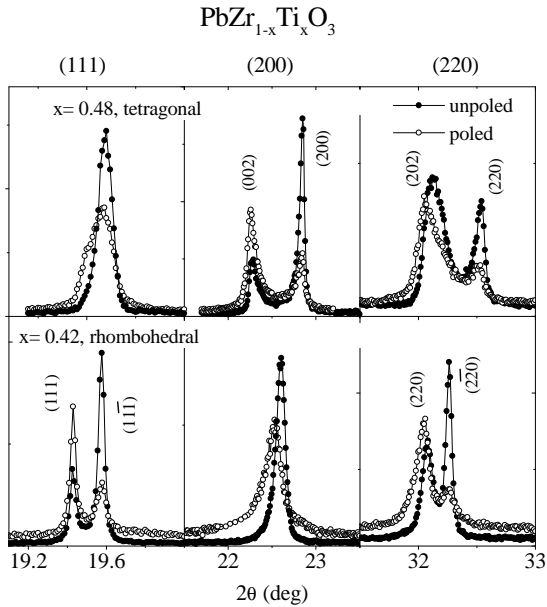}
\caption{(111), (200) and (220) pseudo-cubic reflections for the x= 0.48
(tetragonal) and x= 0.42 (rhombohedral) PZT compositions, before and after
poling. After Guo et al. \protect\cite{Noheda4}}
\label{fig6}
\end{figure}

Diffraction experiments with poled ceramics as well as with PZT ceramics
under electric field applied {\it in-situ} were carried out. This
measurement were taken on the flat plate on symmetric reflection, which
means that only scattering vectors perpendicular to the surfaces are measured 
\cite{Noheda4}. A plot of selected diffraction peaks of poled and unpoled
samples (Fig. \ref{fig6}) shows the expected intensity differences which are
attributed to differences in domain populations after poling, in both the
tetragonal (x= 0.48) and a rhombohedral (x= 0.42) compositions. The behavior
of the peak positions after poling was, however, unexpected. As shown in the same 
figure for the rhombohedral composition, the (hhh)
diffraction peaks, corresponding to the polar direction, were not shifted
after poling. A large shift of the (h00) peak position was observed,
however, which means that the piezoelectric elongation of the unit cell is
not along the polar direction, but along [001]. In a similar way, for the
tetragonal composition (x= 0.48), no shift was observed along the polar
[001] direction, while clear poling effects were evident in the (hhh) peaks.
The explanation of this striking behavior lies in the monoclinic phase. The
piezoelectric strain occurs, for compositions close to the MPB not along the
polar axes but along the directions that induce the monoclinic distortion.

All these observations lead us to propose that the so-called morphotropic phase
boundary is not a boundary but rather a phase with monoclinic symmetry. This
new phase is intermediate between the tetragonal and rhombohedral PZT
phases. Its symmetry relates both phases (Cm being a subgroup of both P4mm
and R3m) through the only common symmetry element, the mirror plane. Both,
the tetragonal and rhombohedral phases (at least in the proximity of the
MPB) have a local structure different from the long-range one and at low
temperatures a monoclinic long range order is established by the freezing-out of one
of the ''local monoclinic structures'' in both the rhombohedral and the
tetragonal phases. Under the application of an electric field, one of the
locally disordered sites becomes preferred, inducing the monoclinic
distortion. This induced monoclinic phase is stable and remains after the
field is removed.

These results can explain some of the puzzles in PZT, such as the larger
piezoelectric coefficient found in rhombohedral PZT along the tetragonal
direction \cite{Du}. Taking into account the monoclinic phase, very recent {\it ab initio%
} calculations have been able to explain the high piezoelectric response of these materials by 
considering rotations of the polar axis in the monoclinic plane ($d_{15}$) \cite{Bellaiche2}. 
Indications of a phase of
lower symmetry than tetragonal have been found by optical measurement on
single crystals of PZN-PT close to the MPB \cite{Fujishiro}. Something
similar could be true in other ferroelectric systems with similar MPBs as
PMN-PT or some Tungsten-Bronzes.

We thank L. Bellaiche, T. Egami, A.M. Glazer and C. Moure for helpful discussions, B. Jones for
the excellent samples, and A. Langhorn for his technical support during the
field experiments. Financial support by the U.S. Department of Energy under
contract No. DE-AC02-98CH10886 and ONR under project MURI (N00014-
96-1-1173) is also acknowledged.

\end{document}